# Thermophysical and Mechanical Properties Prediction of Rear-earth High-entropy Pyrochlore Based on Deep-learning Potential

Yuxuan Wang, Guoqiang Lan, Huicong Chen, Jun Song[*]

*Department of Mining and Materials Engineering, McGill, University, Montreal, Quebec, H3A 0C5, Canada*

**Abstract**

High-entropy pyrochlore oxides possess ultra-low thermal conductivity and excellent high-temperature phase stability, making them promising candidate for next-generation thermal barrier coating (TBC) materials. However, reliable predictive models for such complex and disordered systems remain challenging. Ab initio methods, although accurate in describing anharmonic phonon-phonon interactions, struggle to capture the strong inherent phonon-disorder scattering in high-entropy systems. Moreover, the limited simulation cell size, hundreds of atoms, cannot fully represent the configurational complexity of high-entropy phases. On the other hand, classical molecular dynamics (MD) simulations lack accurate and transferable interatomic potentials, particularly in multi-component systems like high-entropy ceramics. In this work, we employed Deep Potential Molecular Dynamics (DPMD) to predict the thermophysical and mechanical properties of rare-earth high-entropy pyrochlore oxide system. The deep-potential (DP) model is trained on a limited dataset from ab initio molecular dynamics (AIMD) calculations, enabling large-scale molecular dynamics simulations with on-the-fly potential evaluations. This model not only achieves high accuracy in reproducing ab initio results but also demonstrates strong generalizability, making it applicable to medium-entropy ceramics containing the same constituent elements. Our study successfully develops a deep potential model for rare-earth pyrochlore systems and demonstrates that the deep-learning-based potential method offers a powerful computational approach for designing high-entropy TBC materials.

---

[*] Author to whom correspondence should be addressed. E-mail address: jun.song2@mcgill.ca (J. Song).

# 1. Introduction

High-entropy pyrochlore oxides [1-8] (HEPOs) have garnered significant research attention due to their exceptional thermal stability and ultra-low thermal conductivity, making them promising candidates for next-generation thermal barrier coatings (TBCs). Compared to conventional single-component ceramics, high-entropy ceramics exhibit disordered chemical composition and distorted lattice structure, which profoundly influence phonon transport, mechanical behavior, and phase stability. HEPOs have been reported to maintain phase stability above 1600 °C [9] while exhibiting thermal conductivity lower than 1.5 W/m·K [10], rendering them hot spots for high-performance thermal protection applications.

The challenge of developing new high-entropy ceramics primarily arises from the vast number of possible elemental combinations, making materials screening a complex and time-consuming process. Traditional experimental approaches require extensive synthesis and characterization efforts, leading to high costs, significant material waste, and prolonged development cycles. As a result, computational simulations have emerged as a powerful tool to accelerate the discovery and optimization of high-entropy ceramics. However, traditional ab initio methods and molecular dynamics (MD) simulations face several challenges. Ab initio calculations, particularly those based on density functional theory (DFT), face several challenges when applied to high-entropy systems. One major challenge is the limitation of cell size: conventional DFT calculations typically accommodate within a few hundred atoms, which is insufficient to capture the configurational complexity inherent in high-entropy materials. As a result, it is difficult to make practical and accurate predictions of their physical properties. Furthermore, unlike conventional single component materials where thermal conductivity is primarily from anharmonic phonon-phonon scattering, high-entropy ceramics exhibit pronounced phonon-disorder scattering. However, effective prediction models that simultaneously contain both scattering mechanism remain undeveloped. In addition, static DFT calculations are performed at 0 K and thus fail to incorporate temperature effects, which are crucial for evaluating functional properties. Although ab initio molecular dynamics (AIMD) provides a realistic framework for simulating temperature-dependent behaviors, it is computationally prohibitive. For instance, simulating systems with only a few hundred atoms already imposes a substantial computational burden, yet this size remains inadequate to fully capture the effects of high configurational entropy.

Compared to ab initio methods, molecular dynamic (MD) that based on Newton's mechanics has been widely used in the study of high-entropy materials due to their relatively low computational cost and larger scale [11-14]. However, its accuracy heavily depends on the quality of the interatomic potential, which presents significant challenges in modeling the complex interactions within high-entropy ceramics. The chemically disordered nature of these materials results in intricate local atomic environments and influences both short-range and long-range interactions [15]. Conventional empirical potentials often struggle to accurately describe these effects, as they are typically parameterized for specific structures and lack transferability to high-entropy compositions.

Deep-learning Potentials (DLPs) [16] are recently developed machine-learning-based interatomic potentials designed to accurately model atomic interactions with near first-principles accuracy while maintaining computational efficiency for large-scale MD simulations. Unlike traditional empirical or semi-empirical potentials, DLPs utilize deep neural networks (DNNs) to learn complex potential energy surfaces (PES) from ab initio datasets, such as DFT calculations. As a result, DLPs have emerged as a powerful tool for studying high-entropy materials. In recent years, several relative studies have successfully applied to investigate various properties of high-entropy materials, demonstrating their potential in this field. Kostiuchenko et al. [17]. employed machine-learning potentials combined with Monte Carlo simulations to investigate phase transitions in NbMoTaW HEAs, revealing that local relaxation significantly stabilizes the single-phase bcc structure at room temperature. Pikalova et al. [18] utilized artificial neural network (ANN) potentials trained on DFT data to predict the mechanical properties of high-entropy carbides $(Ti_{0.2}Zr_{0.2}Hf_{0.2}Nb_{0.2}Ta_{0.2})C$, successfully reproducing elastic coefficients, bulk modulus, and Poisson's ratio via MD simulations. Dai et al.[19, 20] applied DLPs to study the temperature-dependent thermal and elastic properties of high-entropy borides $(Ti_{0.2}Zr_{0.2}Hf_{0.2}Nb_{0.2}Ta_{0.2})B_2$ and carbides $(Zr_{0.2}Hf_{0.2}Ti_{0.2}Nb_{0.2}Ta_{0.2})C$, accurately capturing anisotropic thermal expansion, phonon thermal conductivity, and elastic moduli up to 2400°C. Balyakin et al. [21] used an ANN-based interatomic potential to calculate the melting point of CoCrFeMnNi, employing a two-phase method within MD simulations, and obtained results consistent with experimental data. Collectively, these studies highlight the ability of DLPs to model complex high-entropy systems with near first-principles accuracy while maintaining computational efficiency for large-scale simulations. However, relative potential model haven't been built and studied within HEPOs system.

In this work, we studied the availability of using DLPs to predict thermophysical and mechanical properties for HEPOs. By training a Deep Potential (DP) model [16, 22] on a limited set of AIMD dataset, we realized large-scale molecular dynamics simulations with on-the-fly potential prediction, ensuring both computational efficiency and high accuracy. Compared to conventional MD approaches, our model exhibits strong transferability, making it applicable to medium-entropy ceramics within same compositional space. The results demonstrate that this well-trained DP model effectively captures interatomic interactions in complex high-entropy pyrochlore systems, with predictions of thermophysical and mechanical properties aligning well with experimental data.

## 2. Methodology

Fig. 1(a) illustrates the workflow of utilizing the Deep Potential (DP) model for molecular dynamics (MD) simulations, encompassing AIMD data generation, model training, and dynamic simulations. Initially, ab initio molecular dynamics (AIMD) simulations were conducted with limited simulation time and small supercell sizes. The generated dataset, including atomic coordinates $R_i(x_i, y_i, z_i)$, interatomic forces $F_{i,j}$, total energy $E_{tot}$, and atomic types, were then used for model training. The DP model, comprising an embedding network and a fitting network, is trained in a GPU environment. Once trained, the DP model is integrated into MD simulation software (DPMD), enabling large-scale and long-time investigations of the mechanical and thermophysical properties of pyrochlore oxides. Details are described below.

### 2.1 Ab initio molecular dynamic simulations

The training dataset for the DP model was generated using AIMD simulations performed with the Vienna Ab initio Simulation Package (VASP)[23]. Both single-component and high-entropy pyrochlore supercells were constructed to ensure comprehensive configurational sampling. For the single-component pyrochlore structures, supercells of $La_2Zr_2O_7$, $Nd_2Zr_2O_7$, $Sm_2Zr_2O_7$, $Eu_2Zr_2O_7$, $Gd_2Zr_2O_7$ and $Yb_2Zr_2O_7$ were generated by expanding the primitive unit cell into a $2 \times 2 \times 2$ supercell, comprising 176 atoms. For the high-entropy pyrochlore compositions, the special quasi-random structure[24] (SQS) method was employed to construct the same size of supercells of following compositions: $((La_{0.2}Sm_{0.2}Eu_{0.2}Gd_{0.2}Yb_{0.2})_2Zr_2O_7$ (LSEGY), $(La_{0.2}Nd_{0.2}Eu_{0.2}Gd_{0.2}Yb_{0.2})_2Zr_2O_7$ (LNEGY) and $(La_{0.2}Nd_{0.2}Sm_{0.2}Gd_{0.2}Yb_{0.2})_2Zr_2O_7$ (LNSGY). To cover as many as random configuration, ten distinct random supercells were generated for each high-entropy composition. AIMD simulations were subsequently conducted for all supercells, including six single-component and thirty high-entropy configurations, under the NPT ensemble at seven temperature points: 50 K, 300 K, 600 K, 900 K, 1200 K, 1500 K, and 1800 K and pressure of 0 kbar. Each simulation was performed for 1 ps with a timestep of 1 fs to ensure adequate statistical sampling.

For pseudo-potential, generalized gradient approximation (GGA) with the Perdew-Burke-Ernzerhof (PBE) exchange-correlation functional [25] was employed. The electron - core interactions were described using the projector augmented wave (PAW) method with the frozen-core approximation [26]. The valence electron configurations of the constituent elements were defined as follows: La - $5s^25p^66s^25d^1$, Nd - $5s^25p^64f^16s^2$ ($4f^3$ frozen in the core), Sm - $5s^25p^64f^16s^2$ ($4f^5$ frozen in the core), Eu - $5s^25p^64f^16s^2$ ($4f^6$ frozen in the core), Gd - $4f^65d^16s^2$ ($4f^1$ frozen in the core), Yb - $5s^25p^64f^16s^2$ ($4f^{13}$ frozen in the core), Zr - $4s^24p^65s^24d^2$ and O - $2s^22p^4$. The calculations utilized a plane-wave energy cutoff of 520 eV and a Γ-centered k-point mesh of $1 \times 1 \times 1$ for Brillouin-zone integration. And the electronic convergence criterion was set to an energy threshold of $5 \times 10^{-5}$ eV/atom.

## 2.2 Deep-learning potential training

The DP model for rare-earth pyrochlore was trained using the *DeepMD-kit* [16, 22] package, incorporating the DPA-1 descriptor-a novel attention-based mechanism designed to effectively capture the conformational and chemical complexities of atomic systems. The embedding network consists of three layers with 25, 50, and 100 nodes, respectively, while the fitting network comprises three fully connected layers, each containing 200 nodes. The cutoff distances were set to (9.0 Å, 11.0 Å). To ensure robust model training, 80% of the AIMD dataset was randomly selected for training, while the remaining 20% was evenly split into validation and testing datasets.

Fig. 1(b) presents the state space of temperature and pressure in the training data of $La_2Zr_2O_7$, where every dot represents one AIMD-calculated structure frame. It's evident that the training data cover the temperature from around 0 K to 2000 K and pressure between -50 kbar and 50 kbar, from which the DP model will be well-trained for further molecular dynamic simulation.

## 2.3 Molecular dynamic simulations

To enable deep potential molecular dynamic (DPMD) prediction, the *DeePMD-kit* plugin was integrated into the GPU version of LAMMPS [27]. During the simulation, interatomic potential will be real-time calculated through well-trained PD model.

For lattice constant calculations, the supercells were constructed by expanding the conventional unit cell into a 3 × 3 × 3 supercell, consisting of 2376 atoms. During the simulation, the ensemble was set as NPT, where pressure $P$ was set as 0 kbar and temperature $T$ increments followed a stepwise process. The starting and ending temperatures were set to 100 K and 1600 K, respectively. Initially, a 5 ps simulation was performed at 100 K to allow lattice relaxation and achieve equilibrium. Subsequently, the temperature increased by 20 K over 2 ps to reach 120 K, followed by an 8 ps simulation at 120 K to maintain equilibrium. The last 2 ps at 120 K was used to calculate the lattice constant. This process was then repeated incrementally for the entire temperature range.

For thermal conductivity calculation, the non-equilibrium molecular dynamics (NEMD) method was adopted. The supercells were constructed by expanding conventional unit cell into 1 × 1 × 25 (2200 atoms), the length of which is long enough to exclude the size effect. Initially, the system was equilibrated using the NPT ensemble for 10 ns at target temperatures, ranging from 300 K to 1500 K in 100 K increments. Subsequently, the simulation was switched to the NVE ensemble, where a hot region and a cold region were defined, and an equal amount of heat was added and removed, establishing a stable heat flux within the system. After 40 ps of simulation, the temperature distribution reached a steady state. Finally, through Fourier's low:

$$\kappa = -\frac{J}{2S \cdot \nabla T} \qquad (1)$$

where heat flux $J$ equals endothermic/exothermic power, which is set as 0.5 eV/ps, $S$ is cross-sectional area of supercell and $\nabla T$ is temperature gradient, from which the thermal conductivities can be calculated.

For the mechanical properties calculation, the supercells were set identical to those used in the lattice constant calculation. First, the system was equilibrated in the NPT ensemble for 10 ps, with the temperature maintained at 300 K and pressure at 0 kbar. This was followed by a 3 ps simulation under the NVE ensemble to stabilize the system. Next, trivial normal and shear strains were applied to the simulation box to compute stress tensors. Using Hooke's law, the elastic constants $C_{ij}$ were determined. Finally, the mechanical properties were obtained using the following equations:

$$B = \frac{C_{11} + 2C_{12}}{3} \tag{2}$$

$$\nu = \frac{1}{1 + \frac{C_{11}}{C_{12}}} \tag{3}$$

$$E = \frac{3B(1 - 2\nu)}{1 + \nu} \tag{4}$$

from which Bulk modulus $B$, Poisson's ratio $\nu$, Young's modulus $E$ can be calculated.

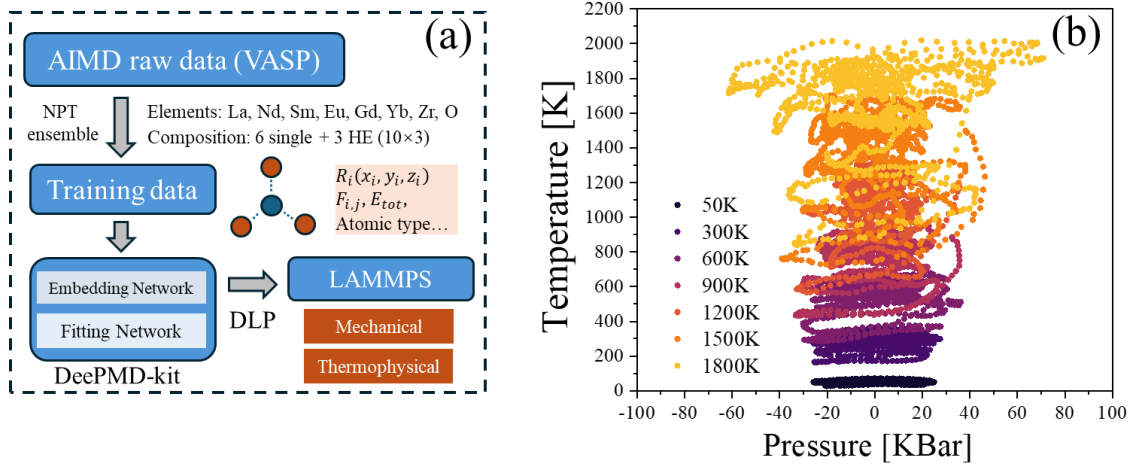

**Fig. 1** (a) Workflow schematic of deep potential molecular dynamic simulation and (b) State space schematic (temperature vs. pressure) of training data set of $La_2Zr_2O_7$.

## 3. Results and discussion

3.1 Model evaluation

To assess the accuracy and predictive capability of the developed model, 10 % of the total DFT calculated dataset was selected for model testing. The model-predicted energies and forces will be compared with that from AIMD. Fig. 2 presents scatter plots of energies and forces for four different compositions: LSEGY, LNEGY, LNSGY, and LEG (($La_{1/3}Eu_{1/3}Gd_{1/3})_2Zr_2O_7$). Notably, while LEG is not part of the training dataset, accurate predictions can still be realized using the trained model. The root mean square error (RMSE) is reported for each case to quantify the deviation between the model predictions and reference DFT values.

Fig. 2(a)-(d) illustrate the relationship between the predicted and AIMD calculated reference energy values. For LSEGY, LNEGY, and LNSGY compositions, the trained models exhibit excellent agreement with AIMD, with RMSE values of $1.002 \times 10^{-3}$ eV/atom, $9.377 \times 10^{-4}$ eV/atom, and $9.118 \times 10^{-4}$ eV/atom, respectively. Among them, LNSGY achieves the lowest RMSE, suggesting that it provides the most accurate energy predictions. Interestingly, the LEG model, despite not being included in the training dataset, also achieves a relatively low RMSE of $1.381 \times 10^{-3}$ eV/atom, indicating that the trained models can generalize well to medium-entropy ceramic within the same constituent space, enabling accurate energy predictions beyond the training set.

Fig. 2(e)-(h) present the force predictions for the same compositions, which again show strong predictive capability. The RMSE values for LSEGY, LNEGY, and LNSGY are $7.682 \times 10^{-2}$ eV/Å, $7.639 \times 10^{-2}$ eV/Å, and $7.554 \times 10^{-2}$ eV/Å, respectively, confirming their reliability in force prediction. Surprisingly, the LEG model achieves an even lower RMSE of $6.313 \times 10^{-2}$ eV/Å, suggesting that, despite not being explicitly included in training, the model still provides highly accurate force predictions for medium-entropy LEG structures. This suggests the model effectively captures the characters of interaction between particle pairs in complex high-entropy system.

The evaluation results confirm that the trained model provide reliable energy and force predictions, with LNSGY performing best in energy accuracy. More importantly, despite LEG not being part of the training set, it can still achieve highly accurate predictions using the trained model, particularly in force calculations. This highlights the strong generalization ability of the developed model and its potential applicability in prediction of unknown composition.

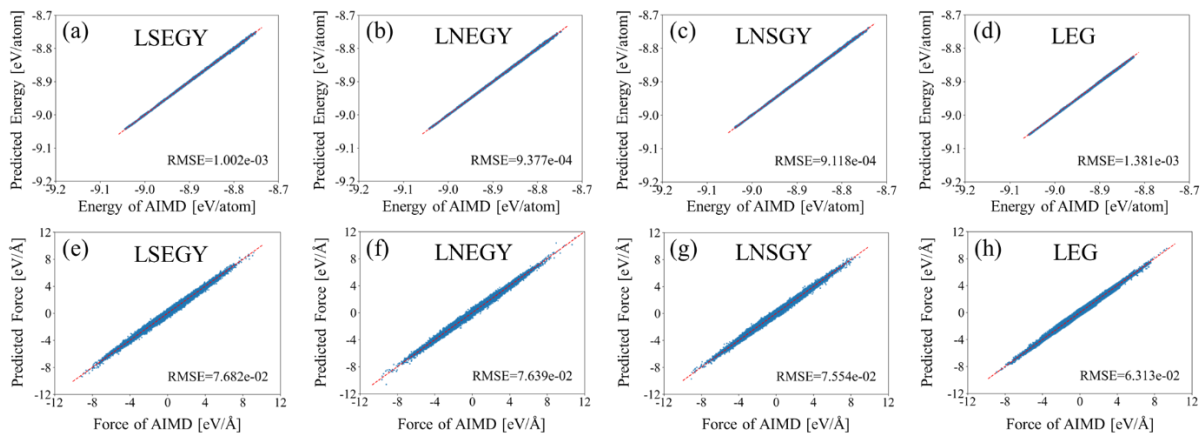

**Fig. 2** Comparison of predicted energy and force values with AIMD reference data for different compositions. (a)-(d) present the correlation between predicted and AIMD calculated energies for LSEGY,

LNEGY, LNSGY, and LEG, respectively, with corresponding RMSE values. (e)-(h) illustrate the force predictions against AIMD calculated forces for the same compositions.

## 3.2 Lattice constant and thermal expansion

Thermal expansion is one of the most important index for TBC design, for that matchable thermal expansion with underlying alloy component will reduce residual stress during thermal cycling and thus prolong service life of the coating [28]. Therefore, to assess the performance of the developed DP model in predicting finite-temperature structural properties, we investigate the temperature dependence of lattice constants and thermal expansion coefficients in different pyrochlore compositions. Fig. 3 presents a comparison between the results obtained from DPMD (solid markers), AIMD simulations (hollow markers), and experimental data [29-31] (half hollow marks). Single component pyrochlore $La_2Zr_2O_7$ (LZO) was also added for benchmark.

Temperature dependence of lattice constants for different pyrochlore compositions were shown in Fig. 3(a). The results obtained from DPMD simulations are represented by solid markers, showing a clear trend of increasing lattice constants with temperature. Among all compositions, $La_2Zr_2O_7$ (LZO) (solid black square markers) exhibits the largest lattice constant across the entire temperature range, followed by LEG, LNSGY, LNEGY, and LSEGY. For LZO, the lattice constantly increases linearly with temperature. However, for LEG, LNSGY, LNEGY, and LSEGY, the lattice constants remain nearly constant between 100 K and 300 K, after which they exhibit a linear increase similar to LZO. This plateau behavior at low temperatures has been identified as a consequence of an anion order-disorder transition in the pyrochlore structure []. Specifically, a fraction of 8b oxygen atoms migrate into 48f vacant sites, leading to lattice contraction. Although the relative lattice constant trend (LZO > LEG > LNSGY > LNEGY > LSEGY) is consistent with experimental observations, the absolute values obtained from DPMD are systematically ~0.1 Å larger than the experimental results (hollow markers). This discrepancy primarily arises from the accuracy limitations of the AIMD training dataset that used to develop the DP model, where the GGA potential will usually overestimate the lattice constant in most of the prediction cases [32-34]. Comparing DPMD (solid markers) with AIMD (hollow markers), a strong agreement is observed across most of the temperature range. However, minor deviations appear in medium- and high-entropy compositions at low temperatures, which likely stem from the finite simulation time of AIMD (~1 ps). The lattice constants may not have fully relaxed, leading to small discrepancies.

Fig. 3(b) presents the temperature dependence of the thermal expansion coefficient (CTE) for various pyrochlore compositions. For DPMD results, the CTE curve is calculated from lattice constant of Fig. 3(a). The CTE of LZO decreases from approximately $13 \times 10^{-6}$ K$^{-1}$ to $11 \times 10^{-6}$ K$^{-1}$ as the temperature increases from 200 K to around 800 K. Beyond 800 K, the CTE remains relatively stable, fluctuating around $11 \times 10^{-6}$ K$^{-1}$. For other medium- and high-entropy compositions, the CTE follows a similar trend. In the temperature range of 200 K to approximately 600 K, the CTE exhibits a rapid increase from below $5 \times 10^{-6}$ K$^{-1}$ to around $13 \times 10^{-6}$ K$^{-1}$. Between 600 K and 1200 K, the CTE remains nearly constant at approximately $13 \times 10^{-6}$ K$^{-1}$. Above 1200 K, a slight decrease to around $12 \times 10^{-6}$ K$^{-1}$ is observed. Comparing the simulation results with experimental data (hollow markers with dashed lines), the CTE obtained from DPMD simulations is consistently higher by approximately $2 \times 10^{-6}$ K$^{-1}$ in the high-temperature range. This discrepancy may arise from the same accuracy limitations of the ab initio molecular dynamics (AIMD) training data. Improving the accuracy of the training dataset could enhance the reliability and applicability of DPMD simulations for predicting thermal expansion behavior.

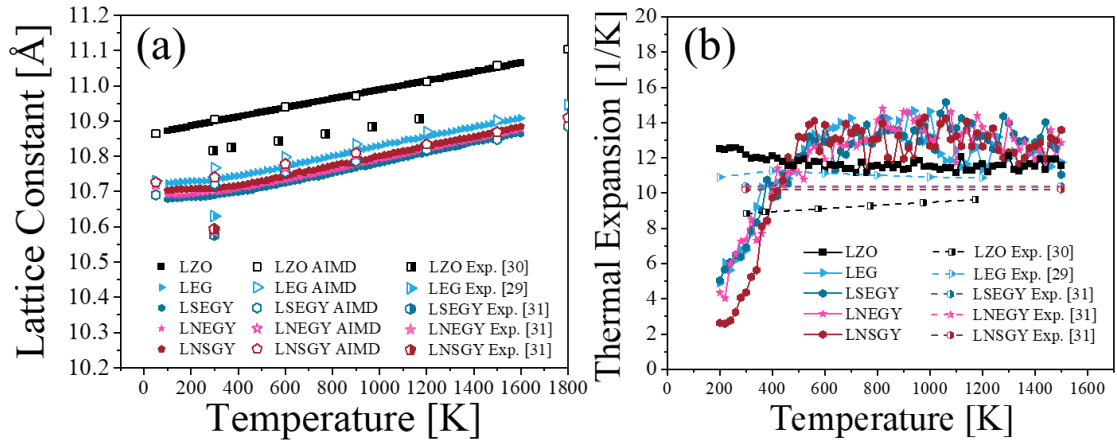

**Fig. 3** Temperature dependence of (a) lattice constants and (b) thermal expansion coefficients of DPMD (solid markers), AIMD (hollow markers) and experimental data[29-31] (half hollow markers) for LZO (La$_2$Zr$_2$O$_7$, black square), LEG (light blue triangle), LSEGY (lark blue hexagon), LNEGY (pink pentagram) and LNEGY (brown pentagon).

## 3.3 Mechanical properties

In thermal barrier coating (TBC) applications, mechanical properties play a crucial role in ensuring coating stability, durability, and resistance to damage[28]. Appropriate mechanical properties enhance thermal resistance and extend the service life of the coating. In this chapter, the mechanical properties of various pyrochlore oxides, including elastic constants ($C_{11}$, $C_{12}$, $C_{44}$) Poisson's ratio, bulk modulus, and Young's modulus, are predicted using DPMD method. The computed Young's modulus values are further compared with available experimental measurements and other theoretical calculations. The results are summarized in Tab. 1.

Among the single-component pyrochlore oxides, $Yb_2Zr_2O_7$ (YZO) exhibits the highest $C_{11}$ value (280.24 GPa), whereas $Sm_2Zr_2O_7$ (SZO) has the lowest (243.71 GPa), indicating that YZO possesses the highest resistance to uniaxial compression. The shear modulus-related $C_{44}$ values are also composition-dependent, decreasing from $La_2Zr_2O_7$ (LZO, 77.13 GPa) to $Yb_2Zr_2O_7$ (46.20 GPa), indicating a trend here increasing cationic substitution with smaller rare-earth elements leads to a reduction in shear rigidity. Poisson's ratio varies from 0.219 (YZO) to 0.271 (LZO), where lower values indicate increased brittleness and a reduced ability to accommodate plastic deformation. The bulk modulus, which characterizes resistance to volumetric compression, also varies, with YZO exhibiting the highest value (144.01 GPa). Regarding Young's modulus, LZO exhibits the lowest value (188.61 GPa), while YZO has the highest (242.80 GPa). These results indicate that a smaller cationic size leads to a higher Young's modulus, a trend that aligns well with experimental data.

For medium- and high-entropy pyrochlore oxides, including LEG, LSEGY, LNEGY, and LNSGY, the same mechanical indices were computed. Compared to high-entropy compositions, the medium-entropy composition (LEG) presents a lower $C_{11}$ value (234.67 GPa) and a higher $C_{44}$ value (70.45 GPa), indicating lower resistance to uniaxial compression but higher resistance to shear stress. Poisson's ratio values for both medium- and high-entropy compositions remain similar. For both bulk modulus and Young's modulus, high-entropy compositions exhibit comparable values, which are slightly higher than those of the medium-entropy composition (LEG).

Compared to experimental results and other first-principles calculations, the DPMD-derived Young's modulus values generally present lower values. For LZO, the DPMD-predicted Young's modulus is 188.61 GPa, which is slightly lower than the experimental value of 199.0 GPa and theoretical calculation of 214.0 GPa. Similarly, for NZO, the DPMD calculation yields 200.45 GPa, which is lower than both the experimental measurement 219.0 GPa and the other computational result of 224.0 GPa. In the case of SZO, the DPMD-predicted Young's modulus is 209.83 GPa, compared to the experimental value of 31.0 GPa and the other theoretical calculation of 226.0 GPa. Same deviation also happens in medium and high entropy compositions.

**Table 1** DPMD-predicted mechanical properties ($C_{11}$, $C_{12}$, $C_{14}$, Possion's ratio, Young's modulus) of pyrochlore oxides system and comparison with experimental and other first-principles results.

| Serial | Composition | $C11$ | $C12$ | $C44$ | Poisson Ratio | Bulk Modulus (GPa) | Young's Modulus (GPa) | Exp. Modulus (GPa) | Other Cal. Modulus (GPa) |
|---|---|---|---|---|---|---|---|---|---|
| LZO | $La_2Zr_2O_7$ | 234.92 | 88.35 | 77.13 | 0.271 | 137.27 | 188.61 | 199.0 [31] | 214.0 [35] |
| NZO | $Nd_2Zr_2O_7$ | 238.15 | 80.05 | 74.36 | 0.248 | 132.44 | 200.45 | 219.0 [36] | 224.0 [33] |
| SZO | $Sm_2Zr_2O_7$ | 243.71 | 74.53 | 69.32 | 0.236 | 132.47 | 209.83 | 231.0 [36] | 226.0 [33] |
| EZO | $Eu_2Zr_2O_7$ | 249.33 | 79.91 | 64.92 | 0.239 | 136.54 | 213.82 | 228.9 [37] | N/A |
| GZO | $Gd_2Zr_2O_7$ | 253.78 | 78.58 | 58.88 | 0.235 | 136.66 | 217.29 | 235.9 [38] | 218.8 [33] |
| YZO | $Yb_2Zr_2O_7$ | 280.24 | 80.05 | 46.20 | 0.219 | 144.01 | 242.80 | 242.0 [39] | N/A |
| LEG | $(La_{1/3}Eu_{1/3}Gd_{1/3})_2Zr_2O_7$ | 234.67 | 73.25 | 70.45 | 0.238 | 126.84 | 199.39 | 237.9 [29] | N/A |
| LSEGY | $(La_{1/5}Sm_{1/5}Eu_{1/5}Gd_{1/5}Yb_{1/5})_2Zr_2O_7$ | 241.73 | 74.00 | 57.97 | 0.237 | 130.39 | 205.76 | 239.1 [31] | N/A |
| LNEGY | $(La_{1/5}Nd_{1/5}Eu_{1/5}Gd_{1/5}Yb_{1/5})_2Zr_2O_7$ | 242.48 | 75.44 | 55.87 | 0.238 | 130.45 | 205.07 | 211.7 [31] | N/A |
| LNSGY | $(La_{1/5}Nd_{1/5}Sm_{1/5}Gd_{1/5}Yb_{1/5})_2Zr_2O_7$ | 239.05 | 73.68 | 60.25 | 0.238 | 128.96 | 202.73 | 225.6 [31] | N/A |

Fig. 4 plotted the column relation of DPMD predicted Bulk modules and Young's modules among different pyrochlore compositions, from which a more intuitive comparation can be observed. For single component pyrochlore oxides, Young' modules present an apparently increasing trend with cations size while bulk modules have trivial discrepancies. For medium- and high-entropy composition, both the bulk modules of which is slightly lower than that of single component. And for Young's modulus, their values varies between largest value of single component YZO and smallest value of LZO.

The results demonstrate that DPMD is capable of capturing the overall trends of elastic properties in pyrochlore oxides, with reasonably good agreement with experimental and first-principles data. However, the systematically lower Young's modulus values predicted by DPMD suggest potential limitations related to the interatomic potential function, highlighting the need for further refinement and validation.

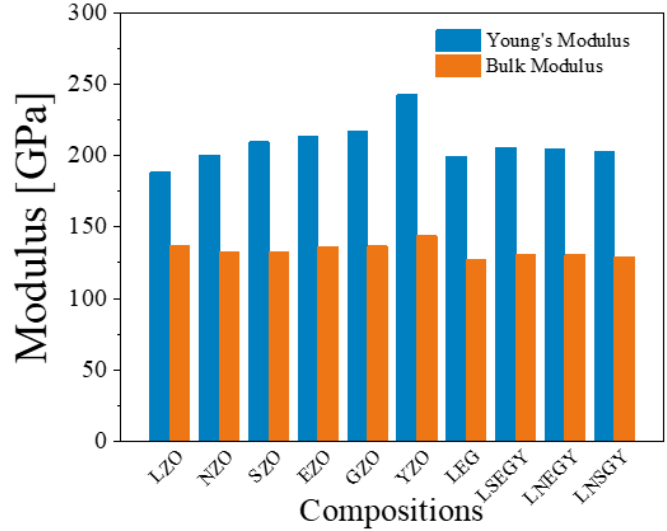

**Fig. 4** Comparison of DPMD-predicted bulk modulus and Young's modulus of varied pyrochlore oxides compositions.

### 3.4 Thermal conductivities

Thermal conductivity is a critical performance metric for thermal barrier coatings (TBCs). Lower thermal conductivity effectively reduces heat flux from the external environment, thereby enhancing the service temperature of coated components [28]. Higher operating temperatures allow gas turbine engines and other high-temperature components to fully utilize their performance potential, leading to improved efficiency and power output. However, traditional predictive models often fail to accurately describe phonon-disorder scattering effects, limiting their ability to provide precise thermal conductivity estimations. To address this limitation, our DPMD model provides a more intuitive and direct approach, enabling a more reliable and physically representative description of thermal transport in disordered ceramic systems.

Fig. 5 presents the thermal conductivity prediction results obtained from DPMD, along with experimental measurements for comparison. Fig. 5(a) illustrates a schematic representation of the NEMD method used to measure the thermal conductivity of LZO. In this approach, a fixed heat flux is applied along the heat conduction direction until a steady-state temperature gradient is established. The thermal conductivity is then calculated from the temperature gradient. The supercell size along the heat flux direction is set to approximately 26 nm, which is longer than most of the phonon mean free path, thus sufficient to exclude the size effect [40].

Figures 5(b)-(f) show the DPMD-predicted thermal conductivity curves as a function of temperature for LZO, LEG, LSEGY, LNEGY, and LNSGY, respectively. Experimental data are included for model evaluation. To better represent the trend of the thermal conductivity curves, a simple empirical relationship, $y=a/T+b$, was used for curve fitting via the least squares method, where $a$ represents the contribution from anhamonic phonon-phonon scattering while $b$ denotes that from other mechanism. The thermal conductivity of all compositions were well fitted using this equation.

For the single-component pyrochlore oxide LZO (c.f. Fig. 5(b)), the thermal conductivity decreases with increasing temperature, which is consistent with Zhang's experimental results[41]. However, Luo's measurements [31] exhibit systematically higher values across all temperature ranges. This discrepancy may arise from differences in sample preparation methods and testing environments, such as variations in grain size, porosity, or measurement techniques.

For the medium-entropy composition LEG (c.f Fig. 5(c)), although this composition was not included in the training dataset, the well-trained model still produces predictions that closely match experimental results. A good agreement is observed between predicted and experimental values in the 300-800 K temperature range. However, at higher temperatures, the experimental results exhibit an abnormal increase, deviating from the expected trend. This anomaly might be attributed to experimental conditions that failed to exclude contributions from photon (radiative) conduction, which becomes significant at elevated temperatures [42].

For the high-entropy compositions (c.f. Fig. 5(d–f)), the thermal conductivity curves show similar trends across different compositions. Compared to single-component LZO, the high-entropy pyrochlore oxides exhibit flatter thermal conductivity curves at lower temperatures, which is reflected in smaller fitted parameter $a$. In the low-temperature regime, high-entropy compositions demonstrate significantly lower thermal conductivity, whereas at higher temperatures, their values become comparable to those of LZO. At lower temperatures, low-frequency phonons dominate heat transport. The observed reduction in thermal conductivity for high-entropy pyrochlore oxides suggests that they induce stronger scattering of low-frequency (long-wavelength) phonons, thereby reducing the overall thermal transport efficiency. When comparing DPMD-predicted results with experimental data, a systematic deviation is observed. Below 1100 K, the DPMD-predicted thermal conductivity values are lower than experimental measurements. Above 1100 K, DPMD-predicted values become higher than experimental results.

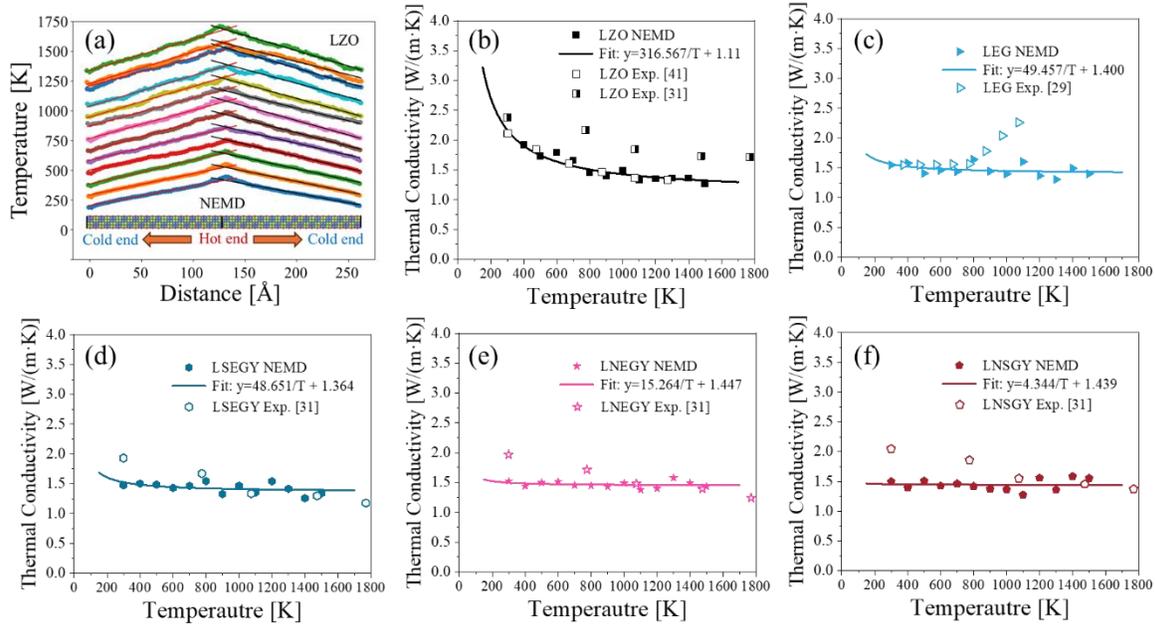

**Fig. 5** DPMD predicted thermal condutivity and comparation with experimental results [29, 31, 41]. (a) Schematic of NEMD method and temperature curves of LZO supercell. (b)-(f) Thermal conductivity curves of LZO, LEG, LSEGY, LNEGY and LNSGY.

Fig. 6 presents the thermal conduction performance comparison for different pyrochlore oxides at 300 K. Fig. 6(a) illustrates the DPMD-predicted thermal conductivity, where LZO exhibits the highest thermal conductivity (~2.1 W/m·K), significantly higher than other compositions, suggesting relatively weaker phonon scattering. In contrast, medium- and high-entropy pyrochlore oxides (LEG, LSEGY, LNEGY, and LNSGY) exhibit much lower thermal conductivity (~1.5 W/m·K), indicating that increased chemical disorder enhances phonon scattering and suppresses heat transport. Fig. 6(b) displays the mechanical-to-thermal conductivity ratio ($E/\kappa$), which serves as an indicator of the phonon scattering rate[43]. Medium- and high-entropy compositions exhibit significantly higher $E/\kappa$ values, suggesting that it is entropy that help to reduces phonon scattering lifetime rather than phonon velocities. This result confirms that medium- and high-entropy pyrochlore oxides offer a superior trade-off between mechanical durability and thermal insulation, making them promising candidates for high-performance TBC applications.

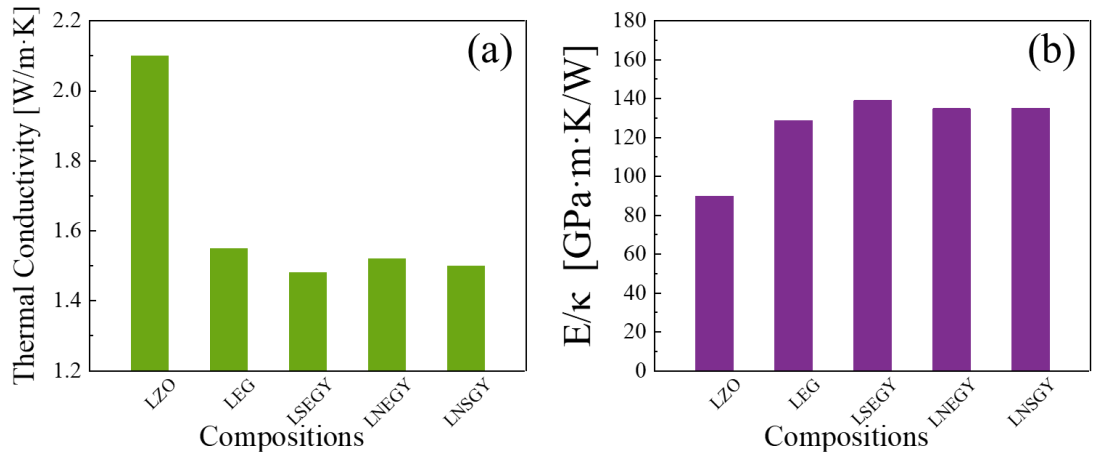

**Fig. 6** Thermal conduction performance comparation for different pyrochlore oxides. (a) DPMD-predicted thermal conductivity at 300K. (b) Calculated E/κ values.

From the above discussion, the DPMD approach successfully captures the thermal conductivity trends in pyrochlore oxides, both in single component, medium- and high entropy compositions, demonstrating good agreement with experimental results in most cases. Medium- and High-entropy compositions exhibit lower thermal conductivity at lower temperatures, supporting the hypothesis that phonon scattering is enhanced by increased chemical disorder. Although discrepancies exist between DPMD predictions and experimental values at high temperatures, they can be attributed to experimental uncertainties and additional heat transfer mechanisms. These results highlight the potential of DPMD as a powerful tool for predicting thermal transport properties in complex ceramic materials, providing new insights into the thermal behavior of high-entropy oxides.

## 4. Conclusion

In this study, a DP model for high-entropy pyrochlore was successful built and examined. By training a DP model on a limited set of AIMD simulations, we demonstrated the capability of machine-learning-based interatomic potentials to capture the complex behavior of chemically disordered ceramic systems with high accuracy. The trained DP model enabled large-scale molecular dynamics simulations while maintaining near first-principles accuracy, providing valuable insights into the structure-property relationships of HEPOs.

By comparing the predicted results with experimental references, our method demonstrates high accuracy and strong generalizability in predicting the thermophysical and mechanical properties of medium- and high-entropy pyrochlore oxides. The results show that DPMD successfully reproduces thermal conductivity trends, accurately describing phonon scattering effects induced by disorder system. Additionally, DPMD-predicted elastic properties, including elastic constants, bulk modulus, and Young's modulus, exhibit reasonable agreement with experimental and first-principles calculations, further validating its reliability. Although minor discrepancies exist, they are likely attributed to experimental uncertainties and potential limitations of the training dataset.

Overall, this work demonstrates that Deep-Potential Molecular Dynamics is a powerful and scalable computational tool for studying the thermophysical and mechanical properties of high-entropy ceramics. Future research should focus on expanding the constituent space and further refining the DP model, especially the accuracy of training dataset, ultimately advancing the computational discovery and design of novel high-entropy ceramics for thermal barrier coating materials.

## CRediT authorship contribution statement

**Yuxuan Wang:** Conceptualization, Data curation, Methodology, Formal analysis, Validation, Visualization, Writing - original draft, Writing - review & editing; Interpretation of Results. **Huicong Chen:** Methodology, Formal analysis; **Guoqiang Lan:** Formal analysis; **Jun Song:** Supervision, Funding acquisition, Interpretation of Results, Writing - review & editing.

## Declaration of Competing Interest

The authors declare that they have no known competing financial interests or personal relationships that could have appeared to influence the work reported in this paper.

## Acknowledgements

The author acknowledges financial support by National Science and Engineering Research Council of Canada (Grant #: NSERC RGPIN-2023-03628) and McGill Engineering International Tuition Award (MEITA), and the Digital Research Alliance of Canada for providing computing resources.